# Combining randomized and non-randomized data to predict heterogeneous effects of competing treatments

Konstantina Chalkou[1,2], Tasnim Hamza[1,2], Pascal Benkert[3], Jens Kuhle[4,5], Chiara Zecca[6,7], Gabrielle Simoneau[8], Fabio Pellegrini[9], Andrea Manca[10], Matthias Egger[1,11], Georgia Salanti[1]

**Affiliation:** Institute of Social and Preventive Medicine, University of Bern, Bern, Switzerland[1]; Graduate School for Health Sciences, University of Bern, Switzerland[2]; Department of Clinical Research, University Hospital Basel, University of Basel, Basel, Switzerland[3]; Multiple Sclerosis Centre, Neurologic Clinic and Policlinic, Departments of Head, Spine and Neuromedicine, Biomedicine and Clinical Research, University Hospital Basel and University of Basel, Basel, Switzerland[4]; Research Center for Clinical Neuroimmunology and Neuroscience (RC2NB), University Hospital and University of Basel, Basel, Switzerland[5]; Multiple Sclerosis Center, Neurocenter of Southern Switzerland, EOC, Lugano, Switzerland[6]; Faculty of biomedical Sciences, Università della Svizzera Italiana, Lugano, Switzerland[7]; Biogen Canada, Toronto, ON, Canada[8]; Biogen Digital Health, Biogen Spain, Madrid, Spain[9]; Centre for Health Economics, University of York, York, UK[10]; Population Health Sciences, Bristol Medical School, University of Bristol, Bristol, UK[11]

<u>Correspondence to:</u>

Konstantina Chalkou, MSc

Institute of Social & Preventive Medicine

University of Bern

Mittelstrasse 43, 3012 Bern, Switzerland

konstantina.chalkou@ispm.unibe.ch

*Abstract*

Some patients benefit from a treatment while others may do so less or do not benefit at all. We have previously developed a two-stage network meta-regression prediction model that synthesized randomized trials and evaluates how treatment effects vary across patient characteristics. In this article, we extended this model to combine different sources of types in different formats: aggregate data (AD) and individual participant data (IPD) from randomized and non-randomized evidence. In the first stage, a prognostic model is developed to predict the baseline risk of the outcome using a large cohort study. In the second stage, we recalibrated this prognostic model to improve our predictions for patients enrolled in randomized trials. In the third stage, we used the baseline risk as effect modifier in a network meta-regression model combining AD, IPD RCTs to estimate heterogeneous treatment effects. We illustrated the approach in the re-analysis of a network of studies comparing three drugs for relapsing-remitting multiple sclerosis. Several patient characteristics influence the baseline risk of relapse, which in turn modifies the effect of the drugs. The proposed model makes personalized predictions for health outcomes under several treatment options and encompasses all relevant randomized and non-randomized evidence.

Abstract: 210 words

*1 Introduction*

Applications of network meta-analysis to health care questions typically report population-average results about the effects of competing treatments.[1,2] The applicability of such results is limited for decision-making purposes, as some patients might benefit greatly from a treatment while others may do so less or do not benefit at all. Network meta-regression models of studies with individual patient data (IPD) can be used to estimate such heterogeneous treatments effects and indicate the preferable treatment for each patient based on their characteristics.[3,4,5]

One frequently used method to examine whether treatment effects vary across individuals is to divide patients into subgroups based on possibly relevant baseline characteristics, e.g., by performing separate univariate network meta-regression models by age, gender, etc.[6,7] In practice however there are numerous characteristics that can influence the treatment effects. A series of 'one-variable-at-a-time' subgroup analyses is a suboptimal approach; underpowered and prone to false-positive results because of multiple comparisons, which fails to address jointly all clinically relevant effect modifiers.[6,7] Therefore, network meta-regression models aiming to predict the optimal treatment for each patient should preferably account for all relevant individual's baseline characteristics simultaneously. The most commonly used methods for individualized predictions are the effect modeling and the risk modeling approaches.[8,9,10,11,12,13]

The *effect modeling approach* is a regression model with multiple variables and interaction terms between them and the treatment.[7] The effect modeling approach is a flexible method for predicting individualized treatment effects but presents some practical difficulties.[11,7] First, it is prone to overfitting because often a large number of model parameters must be estimated from a small or insufficient sample size.[14,15,16] Although penalization approaches could potentially alleviate the risk of overfitting, research on penalization in (network) meta-regression models is still at an experimental phase.[12] Risk modeling approaches have been developed as a solution to these shortcomings.

The *risk modelling approach i*s a two-stage method to estimate heterogeneous treatment effects within a trial. The approach assumes that the risk of the outcome estimated at baseline (often a proxy for the severity of the condition, the presence of comorbidities etc.) could moderate the treatment effects. [3,4,5,7,11,17] In the *first stage*, the outcome risk for each patient is predicted according to their characteristics at baseline. In the *second stage*, the interaction between the baseline risk and the treatment effect is estimated.[3,6,11,18,19,20,21] The same trial data (internal risk modelling) or different datasets (external risk modelling) can be used at each stage.[3,4,5,7] The risk modelling approach can be thought of as a parameter reduction method, which reduces the risk of overfitting which is one of the most important problems when dealing with a large number of treatment-covariate interactions. Risk modelling outperforms the effect modification method in terms of dimensionality, power, and limited prior knowledge by taking advantage of well-established penalization and variable selection methods in multivariable prognostic models.[11] In addition, risk modelling method can take advantage of the numerous existing established penalization and variable selection methods in multivariable prognostic models (e.g., as the required one in the first stage). Risk modelling method is already used in clinical practice,[7,11,12] and was in detail described and developed, obtaining information from a single RCT with IPD. [12,19] Analysis of RCTs using risk modelling has been successfully applied in various clinical areas to estimate heterogeneous treatment effects.[7,11,12,19] The internal risk modelling approach was recently extended into a network meta-regression model of RCTs with IPD.[22]

The aim of this paper is to extend the previously developed model in two directions.[22] First, to use observational rather than randomized studies to develop the prognostic model at the first stage. Second, to increase the power and precision of the estimated treatment effects by including studies that report only aggregated data (AD).[23] We implemented the network meta-regression model in a Bayesian framework, and we used it to predict the probability of experiencing at least one relapse within the next two years for three drugs and placebo in patients with relapsing-remitting multiple sclerosis (RRMS).

*2 Motivating example and data*

Multiple sclerosis is an immune-mediated disease of the central nervous system with various subtypes. Its most common subtype is RRMS.[24] Patients with RRMS present with acute or subacute symptoms (relapses) followed by periods of complete or incomplete recovery (remissions).[25] Reduction in relapse rates has been commonly used as the primary efficacy endpoint in phase III randomized clinical trials leading to market approval of drugs.[24] Some of the drugs, like natalizumab, are associated with rare but serious side effects, while others, like dimethyl fumarate, are considered to be safer options.[26,27]

We illustrate the methods in datasets including patients with confirmed RRMS. **Table 1** presents the outcome and the patients' baseline characteristics for the included datasets. The outcome of interest is the number of patients that experience at least a relapse within two years from baseline.

*Observational evidence*

We included 935 patients enrolled in the Swiss Multiple Sclerosis Cohort (SMSC).[28] Each patient contributed one, two or three cycles of two-years of follow-up. At the beginning of each cycle, several patient characteristics were recorded, such as age, gender, Expanded Disability Status Scale (EDSS), etc., and we considered them as 'baseline characteristics' for the respective cycle. We included 1752 patient cycles in total.

*Randomized evidence*

We had access to IPD from three phase-III RCTs with 3590 patients assigned to placebo, natalizumab, dimethyl fumarate or glatiramer acetate.[29,30,31] We included a subset of 2150 patients with complete covariate and outcome information, assuming that any missingness does not depend on the risk of relapsing.

*2 Methods*

We proposed a three-stage model. In the first stage, we built a prognostic model using the SMSC. In the second stage, we recalibrated the model to estimate the baseline risk in patients enrolled in RCTs. In the third stage, we estimated heterogeneous treatment effects from a network meta-regression model that synthesizes AD with IPD from RCTs and includes the baseline risk as a prognostic factor and effect modifier.

All our analyses were done in R [32] version 3.6.2 and in JAGS [33] (called through R). The code can be found in the GitHub repository: https://github.com/htx-r/Reproduce-results-from-papers/tree/master/ThreeStageModelRRMS

*Notation*

Consider a set of treatments $\mathcal{H}$ each denoted by $h = 1, 2, \ldots, T$. Let $y_{ijh}$ denote the dichotomous outcome for individual $i=1, 2, \ldots, n$ under treatment $h$ in the $j^{\text{th}}$ trial, and a total of $ns$ trials. An individual can experience the outcome ($y_{ijh} = 1$) or not ($y_{ijh} = 0$). $PF_{ijk}$ is the $k^{\text{th}}$ prognostic factor, $k = 1, 2, \ldots np$. The $np$ prognostic factors are used to estimate the baseline risk $R_i$ (independent of treatment), for each participant. The probability of the outcome to occur for individual $i$ in study $j$ under treatment $h$ is denoted by $p_{ijh}$ and depends on treatment, baseline risk $R_i$ and the interaction between the baseline risk and the treatment. We use asterisk (*), to differentiate between the estimations before and after re-calibration: $R_i^*$ indicates the baseline risk before the re-calibration, estimated using the SMSC, while $R_i$ indicates the baseline risk after re-calibration, for the RCTs population.

*Stage 1: Development and internal validation of the baseline risk prognostic model*

There is plenty of guidance about how to develop and validate a prognostic model.[34,35,36,37] Good practice involves the use of appropriate model selection methods (or pre-specifying the model), shrinkage in the coefficients to avoid extreme predictions, accounting for missing data and correcting for optimism when the model performance is evaluated internally.

In our approach we developed the prognostic model using a non-randomized study for the baseline risk, $R_i^*$, for each individual $i$. We used a logistic mixed-effects model, which accounts for information about the same patient from different cycles. ($c$, where $c = 1, 2, ..., nc$), in a Bayesian framework as

$$logit(R_i^*) = \beta_0^* + u_{0i}^* + \sum_{k=1}^{np}(\beta_k^* + u_{ki}^*) \times PF_{ik} \qquad (Equation\ 1)$$

$\beta_0^*$, and $\beta_k^*$ are the fixed effect intercept and fixed effect slopes respectively, and $u_{0i}^*$ and $u_{ki}^*$ are the individual-level random effects intercept and individual-level random effects slopes, which account for information about the same patient from different cycles. A detailed description of the model development and internal validation is available elsewhere.[38]

*Stage 2: Re-calibration of the baseline risk prognostic model for populations included in RCTs*

Using observational data in stage 1 might lead to better estimation of the prognostic effect of baseline covariates,[39,40,41,42] but predictions for different populations might be less accurate. Re-estimation of all regression coefficients for a new population (ignoring pre-existing prognostic models developed for other settings) might be an option; however, this approach potentially replaces reliable but slightly biased estimates with unbiased but unreliable estimates. Instead, based on relevant guidelines, existing prognostic models should be re-calibrated to make predictions for different populations—here for RCT populations, who are rather different to those included in SMSC.[43,36] To do so, three methods are recommended in the literature: 1) re-calibration of intercept, 2) re-calibration of intercept and overall calibration slope, and 3) re-calibration of intercept, overall slope and re-estimation of some of the regression coefficients. [43,36]

We considered recalibrating only the intercept to ensure that the average predicted baseline risk is equal to average observed baseline risk in RCTs.[43] The recalibrated baseline risk $R_{ij}$ can be estimated by plugging-in the estimated slopes $\beta_k^*$ from stage 1 (Equation 1) and then re-estimate the intercept $\beta_0$, by fitting

$$logit(R_i) = \beta_{0j} + logit(R_i^*) \quad \textit{(Equation 2)}$$

to the RCTs data. The intercept $\beta_{0j}$ could be assumed exchangeable ($\beta_{0j} \sim N(b_0, \sigma_{b_0}^2)$), or common ($\beta_{0j} = b_0$) across studies.

Another option is to recalibrate the intercept and the overall calibration slope, $\beta_{overall}$.[43] This will also update the overall effect of the prognostic factors for the RCTs setting. We first estimate the uncalibrated predictions $R_i^*$ for the RCT population, and then we estimate the following model

$$logit(R_i) = \beta_{0j} + \beta_{overallj} \times logit(R_i^*), \quad \textit{(Equation 3)}$$

where the intercept and the overall regression coefficient of $logit(R_i^*)$ could be assumed exchangeable ($\beta_{0j} \sim N(b_0, \sigma_{b_0}^2)$, $\beta_{overallj} \sim N(b_{overall}, \sigma_{b_{overall}}^2)$) or common ($\beta_{0j} = b_0, \beta_{overallj} = b_{overall}$) across studies. The recalibrated predicted risk score $R_{ij}$ is obtained from equation 3 after estimating $b_0$ and $b_{overall}$.

A more comprehensive option is to re-calibrate the intercept, the overall slope (as above) and in addition re-estimate some of the regression coefficients as needed.[43] The re-estimated baseline risk for RCT patients, $R_i$, will be finally estimated as:

$$logit(R_i) = \beta_{0j} + \sum_{k=1}^{np} \beta_{kj} \times PF_{ik}, \quad \textit{(Equation 4)}$$

where $\beta_{0j}$ and $\beta_{kj}$ are the recalibrated intercept and regression coefficients, and as before can be assumed exchangeable ($\beta_{0j} \sim N(b_0, \sigma_{b_0}^2)$, $\beta_{kj} \sim N(b_k, \sigma_{b_k}^2)$) or common ($\beta_{0j} = b_0, \beta_{kj} = b_k$) across studies.[43]

The common-effects assumption for $\beta_{0j}, \beta_{overallj},$ and $\beta_{kj}$, ignores trial differences, and could be used only in cases where RCTs were designed similarly, have similar protocols and inclusion criteria. In addition, these parameters could be also estimated independently (e.g., each $\beta_{0j}$, $\beta_{\text{overall} j}$, and $\beta_{kj}$ is given a prior distribution); however, this would lead to different baseline risk scores per study, which in turn, would pose challenges in the baseline risk score estimation of an external dataset (other than the RCTs used for the recalibration).

In the application we use the re-calibration method associated with the best balance between model's calibration (i.e., the agreement between the observed outcome's proportions and the predicted probabilities) and discrimination ability (i.e., Area Under the Curve (AUC)).

*Stage 3: Network meta-regression with individual and aggregate data using the baseline risk as prognostic factor and effect modifier*

In the third stage, we used the calibrated $logit(R_i)$ from stage 2 as covariate in a network meta-regression model.[23] We extended the meta-regression model suggested by Saramago et. al. to combine IPD and AD in a network of trials comparing multiple treatments.[23] *In the first part*, we modelled studies with IPD

$$y_{ijt} \sim Bernoulli(p_{ijh})$$

$$logit(p_{ijh}) = \begin{cases} u_j + g_{0j} \times \left(logit(R_i)\right) \text{ if } h = h_{ref,j} \\ u_j + d_{jh_{ref},jh} + (g_{0j} + g^W_{jh_{ref},jh}) \times \left(logit(R_i)\right) + (g^B_{h_{ref},jh} - g^W_{jh_{ref},jh}) \times \overline{logit(R)}^j, \text{ if } h \neq h_{ref,j} \end{cases}$$

*(Equation 5)*

where $\overline{logit(R)}^j$ is the mean logit baseline risk from all patients in study $j$, and each study $j$ has a reference treatment $h_{ref,j} \in \mathcal{H}$.

The parameters of interest are the relative treatment effects $d_{jh_{ref},jh}$. We estimated independently the nuisance parameters $u_j$ for each study (i.e., the log odds of experiencing the outcome under the study's reference treatment). The coefficients $g_{0j}$ measure the prognostic impact of baseline risk and can be assumed independent, exchangeable ($g_{0j} \sim N(\gamma_0, \sigma^2_{\gamma_0})$), or common ($g_{0j} = \gamma_0$) across studies. The regression coefficients $g^W_{jh_{ref},jh}$ measure how the baseline risk of a patient modifies the treatment effect within each study; they can be combined across studies assuming random $\left(g^W_{jh_{ref},jh} \sim N\left(G^W_{h_{ref},jh}, \sigma^2_{G^W}\right)\right)$ or common $\left(g^W_{jh_{ref},jh} = G^W_{h_{ref},jh}\right)$ effects, where $G^W_{h_{ref},jh} = \gamma^W_h - \gamma^W_{h_{ref},j}$ with $\gamma^W_{ref} = 0$ for an overall reference treatment. Similarly, the between-studies effect modification parameters $g^B_{h_{ref},jh}$ measure how the mean baseline risk of each study modifies the relative treatment effect.

*In the second part,* we synthesize information from studies that report only AD. The likelihood of the observed data in AD studies is

$$r_{jt} \sim Binomial(p_{jh}, n_{jh})$$

where $r_{jh}, n_{jh}, p_{jh}$ denote the number of patients experiencing the outcome of interest, the total number of randomized individuals and the probability of experiencing the outcome, in study $j$ in treatment arm h, respectively.

Then, we model the relative treatments effects using the average study-specific baseline risk $\overline{logit(R)}^j$

$$logit(p_{jh}) = \begin{cases} u_j & , if\ h = h_{ref,j} \\ u_j + d_{jh_{ref},jh} + g^B_{h_{ref},jh} \times \overline{logit(R)}^j, & if\ h \neq h_{ref,j} \end{cases} \quad \text{(Equation 6)}$$

To estimate the average baseline risk $\overline{logit(R)}^j$ we simulated pseudo-IPD using a multivariate normal distribution with mean equal to the study-level covariate values $(\overline{PF_{kJ}})$ as reported in the published trials, and variance covariance matrix between the included covariates as estimated through the available IPDs. In this way, we avoid the aggregation bias which would be induced if we simply plugged-in the mean value of each prognostic factor $\overline{PF_{kJ}}$ in the prediction model from stage 2.

The mean values of some of the prognostic factors might not be reported in the original studies. In that case we used imputations to allow studies with partial information on covariates to be included in the meta-regression model, as previously described by Hemming et al (described in Appendix).[44]

*In the third part,* the relative treatment effects, $d_{jh_{ref},jh}$, can be combined across studies in a random-effect ($d_{jh_{ref},jh} \sim N(D_{h_{ref},jh}, \sigma_D^2)$) or common-effect ($d_{jh_{ref},jh} = D_{h_{ref},jh}$) across studies assuming consistency $D_{h_{ref},jh} = \delta_h - \delta_{h_{ref},j}$ where $\delta_{ref} = 0$. Finally, the consistency equations for the parameters $g^B_{h_{ref},jh}$ are $G^B_{h_{ref},jh} = \gamma^B_h - \gamma^B_{h_{ref},j}$ and $\gamma^B_{ref} = 0$.

The difference between $g^W_{jh_{ref},jh}$ and $g^B_{jh_{ref},jh}$ represents an estimate of ecological bias (i.e., the difference between across-study associations and within-study associations, due to study-level confounding).[45] To ensure we do not introduce bias, the within-study effect modification ($g^W_{jh_{ref},jh}$) is estimated through the IPD studies alone (Equation 5), whereas both IPD and AD studies are used for the between-study effect modification estimation ($g^B_{jh_{ref},jh}$). [45,23]

**2.5 Making treatment-specific outcome predictions**

To predict the probability $p_{i_{new}h}$ of the outcome in a new patient $i_{new}$ under treatment $h$, we first estimate $logit(R_{i_{new}})$ from stage 2, and then, under the common treatment effects assumption, we use the following equation

$$logit(p_{i_{new}h}) = a + \delta_h + (\gamma + \gamma^W_h) \times logit(R_{i_{new}}) + (\gamma^B_h - \gamma^W_h) \times (\overline{logit(R)}) \qquad \text{(Equation 7)}$$

The values for $\delta_h, \gamma^W_h$, and $\gamma^B_h$, are those estimated in the third stage of the network meta-regression prognostic model. What data to use to obtain values for $a$—the logit-probability of the outcome under the reference treatment (placebo, in our example),— $\gamma$— the regression coefficient of $logit(R)$, — and $\overline{logit(R)}$—the mean of logit baseline risk across all individuals— depends on the context within which we plan to make predictions. If our aim to is make predictions for RCT populations, then placebo arms can be used to estimate $a, \gamma$ and all RCT patients to estimate the $\overline{logit(R)}$. If we aim to make predictions in a real-world population, then registry or a cohort data should be used instead. Under the random effects assumption, $logit(p_{i_{newh}})$ would be normally distributed with mean as in Equation 7, and a variance-covariance matrix determined by the variance-covariance matrix of $\delta_h$, and $\gamma_h^W$.

**Figure 1** presents a schematic presentation of the aim, data and parameters of each stage of the approach. Information about the studies used in the example are also presented.

*3 Application: heterogeneous effects of treatments for RRMS*

We developed the prognostic model using the SMSC data.[28] The development of the prognostic model in stage one has been previously published and is implemented as Shiny app in https://cinema.ispm.unibe.ch/shinies/rrms/.[38]

We first selected the prognostic factors via a review of the literature.[46,47,48,49,50,51] We included all prognostic factors that were included in at least two previously published prognostic models. The model includes eight prognostic factors: age, sex, EDSS, prior or current treatment (yes or no), months since last relapse, disease duration, number of relapses in the previous two years, number of gadolinium enhanced lesions. We then fitted a logistic mixed-effects regression model in a Bayesian framework accounting for correlations induced by individuals contributing data to more than one cycle. The SMSC includes 1752 observations from two-years cycles of 935 patients, and 302 of those patients experienced at least one relapse. The full model had 22 degrees of freedom (for 10 predictors with random intercept and slope) and the number of events per variable was 13.7. To shrink the coefficients of the regression and avoid extreme predictions, we used Laplace prior distributions.[52] We used multiple imputation to account for missing covariate data.[53,54] After internal validation, the bootstrap optimism-corrected AUC was 0.65 and the bootstrap optimism-corrected calibration slope 0.91. The calibration plot and the evaluation of the model's clinical usefulness are presented elsewhere.[38] The model's accuracy and clinical performance are overall suggesting a useful prediction model.

We then re-calibrated the prognostic model for the RCT setting (stage 2). All three RCTs used for the re-calibration of the baseline risk model in stage 2 were designed under similar protocols from the same company, sharing similar distribution of baseline characteristics as presented in Table 1. Therefore, we assumed common effects across the three RCTs for estimating the intercept (in Equations 2 to 4), the overall regression coefficient (in Equation 3), as well as the regression coefficient for each prognostic factor (Equation 4). Alternatively, when studies share different designs, protocols, and inclusion criteria random effects across studies should be assumed in stage 2. We assessed the predictions of the developed re-calibrated models in a calibration plot with loess smoother (**Appendix figure 1**). All three models appear to predict accurately the probability to relapse within the next two years. The re-calibration method resulting in the highest AUC (AUC=0.61) was "the re-calibration and selective re-estimation" approach (Equation 4); the other two methods resulted in AUC=0.58. Based on the existing literature, risk models with a low predictive ability (0.6 – 0.65) are often adequate to detect risk-based heterogeneous treatment effects.[19, 55] Therefore, we selected the baseline risk from "the re-calibration and selective re-estimation" method to use in the next stage of the risk modelling approach. **Appendix table 1** presents the re-calibrated regression coefficients for each prognostic factor.

In **Figure 2** we show the distributions of the predicted baseline risk by relapsing status in the populations included in the three RCTs. The overall mean predicted baseline risk was 36.8% (95% Credible Interval (CrI) 36.4% to 37.2%). The overlap in the distributions was large, as reflected by the low AUC. For patients who experienced a relapse, the mean predicted risk was 39.2% (95% CrI 38.5% to 39.8%) whereas for patients who did not, it was 35.4% (95% CrI 34.9% to 35.9%).

The predicted baseline risk in the RCT populations was then used in the network-meta-regression model (equations 5 and 6, stage 3). Because only two AD studies were available, we assumed that $g^W_{jh_{ref,jh}}=g^B_{jh_{ref,jh}}$ to enable model convergence. We also assumed that study-specific relative treatment effects do not have any residual heterogeneity beyond what is already captured by differences in baseline risk. As the heterogeneity variance was not well estimated with five studies, we assumed common relative treatment effects ($d_{jh_{ref,jh}} = D_{h_{ref}jh}$) and common effect modification across studies ($g^W_{jh_{ref,jh}} = G^W_{h_{ref,jh}}$). We also assumed common coefficients for the prognostic effect of the baseline risk ($g_{0j} = \gamma_0$), as all three studies with IPD data were very similar in terms of design and patient characteristics.

None of the 2 AD studies provided information about the number of patients with prior treatment, gadolinium enhanced lesions, and months since last relapse; we performed imputations as described in the Appendix. Then, we created 2 pseudo-IPDs (one for each AD study) via a multivariate normal distribution with mean the reported (or imputed) mean covariate values and variance covariance matrix as estimated via the available IPDs. We estimated the baseline risk for each observation in each of the pseudo-IPD datasets. The estimated mean baseline risk for each study is presented in **Table 1**.

**Table 2** shows the estimated parameters from the network meta-regression model (stage 3). The estimated values of $\gamma_0$ indicate that baseline risk is an important prognostic factor for relapse. We first make predictions for the RCT populations and hence we estimate $a$, and $\gamma$ by synthesizing data from placebo individuals across the three RCTs with IPD and $\overline{logit(R)}$ as the mean of $logit(R_i)$ across all individuals in RCTs with IPD. The treatment effects as a function of the baseline risk are shown in **Figure 3**. In addition, Appendix figure 2 presents the final predictions with their 95% CrIs. Natalizumab gives the lowest probability of relapsing over almost the entire baseline risk range. However, its advantage over dimethyl fumarate for patients with low baseline risk (below 30%, on average) is very small. We also assessed the prediction model's accuracy using the calibration plot with loess smoother (Appendix figure 3). The prediction model predicts accurately the probability of relapsing within the next two years.

Additionally, we performed a sensitivity analysis comparing the final predictions from stage 3 under all three recalibration methods (**Appendix table 2,** and **Appendix figure 4**). All three recalibration methods lead to similar final predictions. On average, natalizumab minimizes the predicted probability of relapsing within the next two years (about 7% to 12% mean absolute difference compared to dimethyl fumarate). For low-risk patients (baseline risk≤30%), the probability to relapse is similar under dimethyl fumarate and natalizumab. For high-risk patients (baseline risk≥50%), natalizumab is the drug that minimizes the predicted probability of relapsing within the next two years (about 13% to 19% mean absolute difference compared to dimethyl fumarate).

To make predictions for the Swiss real-world population we estimate $a$ as the logit-probability of relapse in untreated patients in the SMSC and $\overline{logit(R)}$ as the mean of $logit(R_i)$ across all individuals in the SMSC; $\gamma$ was also estimated using SMSC. The results for the SMSC population are presented in **Appendix figure 5**. Often the patients' baseline disease condition is more severe in RCTs than in observational studies, and hence as expected, the distribution of baseline risk is different between the SMSC and the RCTs. However, the relative ranking of therapies for a given baseline risk does not deviate from this of RCTs (presented in **Figure 3**). As in the RCTs population, the advantage of natalizumab over dimethyl fumarate for patients with low baseline risk is non-existing (Appendix figure 5). An interactive version of Figure 3 and Appendix figure 5 has been implemented in https://cinema.ispm.unibe.ch/shinies/srrms/.

**Table 3** summarizes the information in Figure 3 for patients at baseline risk below 30% (low risk) or more than 50% (high risk). These cut-offs were chosen arbitrarily for illustrative purposes. For high-risk patients (8.5% of patients in RCTs) the risk difference for relapse between natalizumab and dimethyl fumarate is 19% favoring natalizumab. For low-risk patients (25% of patients in RCTs), the risk difference between natalizumab and dimethyl fumarate is 1.4%.

*4 Discussion*

We developed a three-stage network meta-analysis approach, where data from different sources and study designs can be synthesized to make predictions for heterogeneous treatment effects. We exemplified our method by predicting the probability of relapse under three active treatments and placebo in patients with RRMS, we made the code available (https://github.com/htx-r/Reproduce-results-from-papers/tree/master/ThreeStageModelRRMS), and we created an online tool to show the predictions in an interactive way ( https://cinema.ispm.unibe.ch/shinies/srrms/ ).

Central to our work is the risk modelling approach. We preferred the risk modelling method for predicting individualized treatment effects over other available methods; first, we consider the risk of overfitting high in effect modification approach, and, secondly, we expected high risk of false-positive results due to numerous 'one-variable-at-a-time' in univariate network meta-regression models. Risk modelling approach has been originally used in a single randomized trial,[3,6,11,18,19,20,21] and extended to meta-analysis[3,11,56] and more recently to network meta-analysis[22] of randomized trials. However, none of these approaches examined combining and making the best use of all available data sources. Observational studies reflect better the real-world populations and conditions[39,40,41,42] and this is why we used a cohort in the first stage of our approach to develop a model that predicts the baseline risk. In addition, we combined AD and IPD to increase the power and precision of the estimated treatment effects.

The approach has several limitations. It requires that at least one IPD dataset per intervention is available. The access to IPD data entails many challenges and difficulties described in detail elsewhere.[57,58,59] The risk modelling approach assumes that the selected variables adequately capture both prognosis and effect modification. This assumption is difficult to evaluate unless the outcome is well studied, and many prognostic studies exist on the topic, which is rarely the case. We developed the three stages in different models instead into a single Bayesian framework. We used an existing prognostic model for the baseline risk (stage 1), and we developed the re-calibration of the baseline risk model (stage 2) within a frequentist setting to take advantage of the software's re-calibration options. Consequently, uncertainty was not accounted between the different stages and the results from stage three might be over-precise (i.e., we expect that the results in Table 3 will have wider intervals). In addition, the imputation method used, although it allows the use of AD studies even if study-level covariates are missing, may not be the optimal one. Other methods, like advanced multiple imputations techniques for study-level characteristics, may be used.[60] Finally, in the RRMS application, we used common treatment effects model (stage 3) to enable model convergence, because of the small number of studies. This assumption can be relaxed if more studies are available.

The implementation of our approach in the RRMS example shows that several patient characteristics influence the baseline risk of relapse, which in turn modifies the effect of treatments. Natalizumab appears to be the optimal treatment (i.e., minimizes the predicted probability of relapsing) over almost the entire baseline risk range. However, its advantage over dimethyl fumarate for patients with low baseline risk is very small or non-existing. Dimethyl fumarate might be the optimal treatment for these patients, as natalizumab is a drug considered less safe.[26,27] The results are in agreement with those of a recent published work,[22] which used only randomized clinical trials for the individualized predictions of treatment effects for relapses in RRMS. The discrimination of the prediction model for the baseline risk (stage 2) was small (AUC=0.61) but sufficient for our aim. The discriminative ability of the existing prognostic models for relapses in MS is generally low (less than 65%), indicating that relapses might be associated with unknown factors.[22,38,61,49] Second, it has been shown that models with low AUC can still be useful when their predictions are used as potential effect modifiers of treatment.[19,55,62] This was the case in our application where we showed clinically meaningful differences between the interventions for different levels of the baseline risk (Figure 3). Note that the findings of this model, are as expected by practitioners and as described in international guidelines, namely that natalizumab is to be administered as a second line treatment. Prognostic scores from models with low AUC were previously used as effect modifiers:see for example the Thrombolysis in Myocardial Infarction (TIMI) and the CHADS2 risk score; both were powerful in detecting the heterogeneous treatment effects of via a risk modelling approach.[63,64,65,66,67]

The application in the example of RRMS shows the potential of our approach, which can make prediction of individualized treatment effects in RCTs and real-world populations; however, it is not ready for use in clinical practice. Decision-making tools need external validation and need to show evidence about all relevant treatment options, before they are considered for use.[68]

The presented approach offers many opportunities for further development. Several bias-adjusted methods have been proposed to combine real-world data and RCTs for average treatment effects estimation in a meta-analysis framework.[69,70,71] Some of them use the baseline risk to adjust for selection bias in the real-world data in a meta-analysis framework.[71,72,73] The approach we presented could be further extended by using observational data to inform not only the baseline risk in stage 1, but also the relative treatment effects (in stage 3) using appropriate bias-adjusted modelling.[69,70,74] Evidence about treatment cost and safety could be incorporated to extend the model further and to better inform clinical decision-making. Besides, it is possible that other study-level characteristics, like the year of randomization and the risk of bias, may also influence the treatment effects. Such variables can be added to the network meta-regression model, if the number of studies permits. Finally, the implementation of the three stages into a single Bayesian model will allow naturally to incorporate uncertainty from all stages in the final result and avoid spuriously overprecise conclusions.

*5 Conclusion*

The proposed approach combines all relevant evidence sources and can be applied to estimate individualized predictions of treatment effects for any health condition. Consequently, it has the potential to assist clinical practice and decision-making towards treatment recommendations and precision medicine.

*Highlights*

**What is already known**

- Recently, a two-stage model which allows for individualized treatment effects predictions between several competing treatments was developed, using individualized participant data from randomized clinical trials

**What is new**

- We extend this model by combining several data sources, such as observational studies on the top of randomized clinical trials as well as incorporation of aggregate data on the top of individual participant data

**Potential impact for RSM readers outside the authors' field**

- Readers will be able to reproduce the suggested method, in any clinical area, to make individualized predictions of treatment effects between several competing treatments into a network meta-analysis framework, while combining several data sources; the methods are described in detail and the codes used for the illustrative example are publicly available

**Data availability statement:** The data that support the findings of this study were made available from Biogen International GmbH and the Swiss Multiple Sclerosis Cohort (SMSC). Restrictions apply to the availability of these data, which were used under license for this study.

**Funding statement and disclaimer** This research was performed as part of the HTx project. The project has received funding from the European Union's Horizon 2020 research and innovation programme under grant agreement No 825162. This dissemination reflects only the authors' view and the Commission is not responsible for any use that may be made of the information it contains.

**Acknowledgements:** KC, TH, AM and GS are funded by the European Union's Horizon 2020 research and innovation program under grant agreement No 825162. ME was supported by the Swiss National Science Foundation (grant 189498). The authors thank Suvitha Subramanian for her assistance on cleaning the data. The authors would like to thank Suvitha Subramaniam for her support and help regarding the SMSC data cleaning.

The Swiss MS Cohort study received funding from the Swiss MS Society and grant funding from Biogen, Celgene, Merck, Novartis, Roche, and Sanofi.

**Conflicts of interest:** KC, TH, PB, JK, ME, AM, and Georgia Salanti declare that they have no conflict of interest with respect to this paper, JK received speaker fees, research support, travel support, and/or served on advisory boards by Swiss MS Society, Swiss National Research Foundation (320030_189140/1), University of Basel, Progressive MS Alliance, Bayer, Biogen, Celgene, Merck, Novartis, Octave Bioscience, Roche, Sanofi. Gabrielle Simoneau and FP are employees of and hold stocks/stock options in Biogen. Ente Ospedaliero Cantonale (employer) received compensation for Chiara Zecca's speaking activities, consulting fees, or research grants from Almirall, Biogen Idec, Bristol Meyer Squibb, Genzyme, Lundbeck, Merck, Novartis, Teva Pharma, Roche.

*Figures*

*Figure 1 Schematic presentation of each stage's aim, suggested data design ant type, and estimated parameters.*

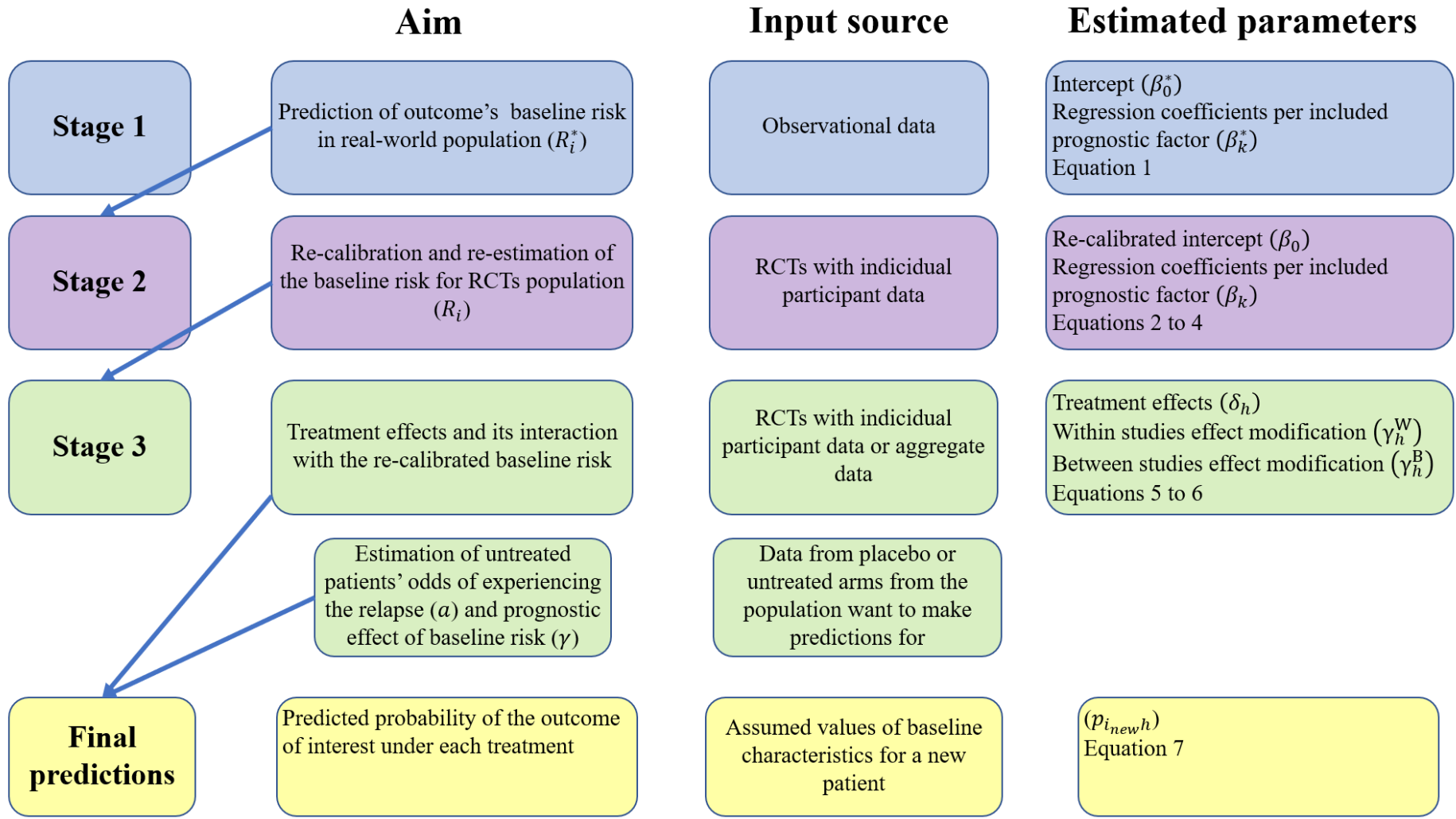

*Figure 2 The distribution of predicted baseline risk of any relapse within the next two years for individuals by relapse status in the RCTs dataset (stage2). The dashed lines indicate the mean of predicted baseline risk for individuals who did experience a relapse (purple) and for those who did not (yellow).*

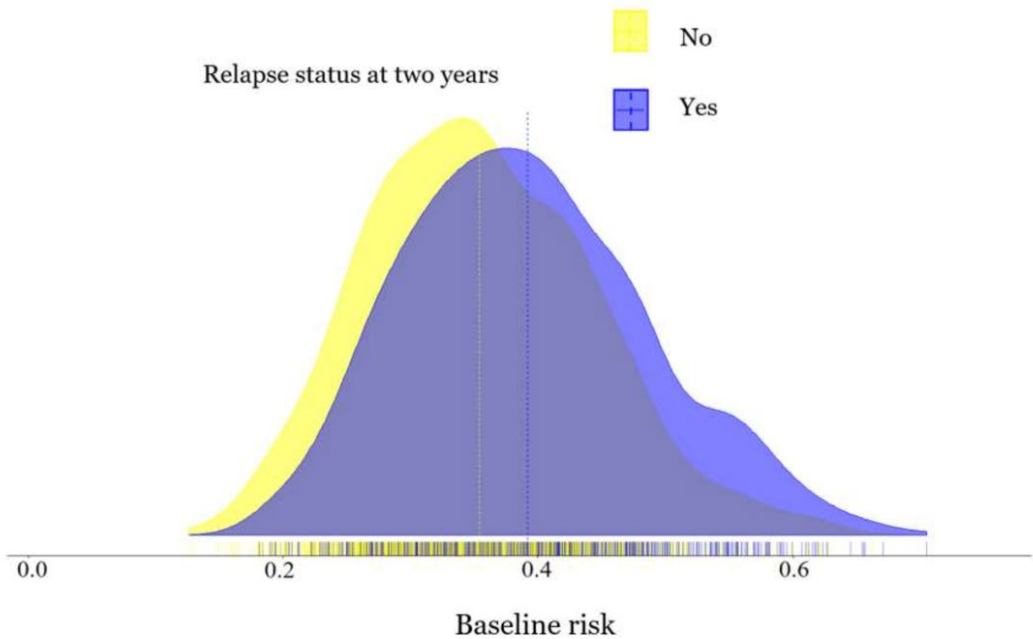

*Figure 3 Predicted probability of relapsing within the next two years as a function of the baseline risk (stage 3) into the randomized clinical trials population. The x-axis shows the baseline risk of relapsing within the next two years (after re-calibration, stage 2) and the y-axis shows the predicted probability to relapse within the next two years under each one of the available treatments. Between the two red vertical dashed lines are the baseline risk values observed in the three randomized clinical trials with individual participant data.[29,30,31] The distribution of the baseline risk in these three trials is presented at the bottom of the graph.*

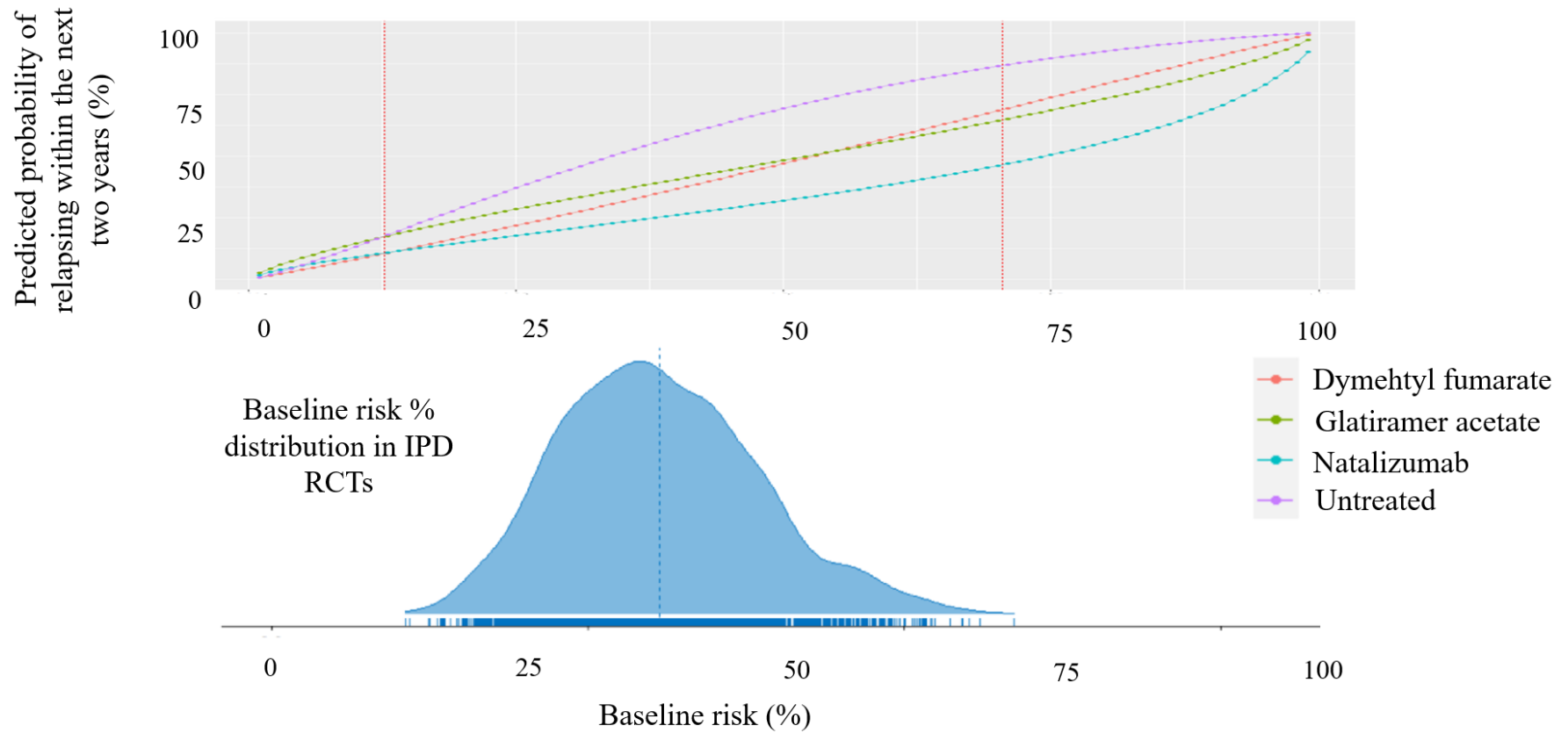

## Tables

**Table 1** Treatment, sample size, outcome, baseline characteristics and baseline risk (stage 2) of patients in the included datasets.

| Study (type of data) | Treatment | Number of patients | Number of patients experiencing relapse at two years (%) | Mean age (sd) | Number of females (%) | Mean baseline EDSS score (sd) | Mean Baseline risk (95%CrI) |
|---|---|---|---|---|---|---|---|
| **SMSC**[19] (real-world study with IPD) | Total | 935 | 191 (20.4) | 40.8 (11.2) | 631 (67.5) | 2.3 (1.4) | 20.1 (2.8, 37.5) |
| **AFFIRM**[20] (RCT with IPD) | Total | 939 | 359 (38.2) | 36.0 (8.3) | 657 (70.0) | 2.3 (1.2) | 36.5 (18.8, 54.1) |
| | Natalizumab | 627 | 183 (29.2) | 35.6 (8.5) | 449 (71.6) | 2.3 (1.16) | 36.9 (19.5, 54.3) |
| | Placebo | 312 | 176 (56.4) | 36.7 (7.8) | 208 (66.7) | 2.3 (1.19) | 35.6 (17.6, 53.7) |
| **CONFIRM**[21] (RCT with IPD) | Total | 1417 | 451 (31.8) | 37.3 (9.3) | 993 (70.1) | 2.6 (1.2) | 37.2 (18.6, 55.7) |
| | Dimethyl fumarate | 703 | 185 (26.3) | 37.8 (9.4) | 495 (70.4) | 2.5 (1.2) | 36.8 (18.2, 55.3) |
| | Glatiramer acetate | 351 | 117 (33.3) | 36.7 (9.1) | 247 (70.3) | 2.6 (1.2) | 37.4 (17.6, 57.3) |
| | Placebo | 363 | 149 (41.0) | 36.9 (9.2) | 251 (69.1) | 2.6 (1.2) | 37.7 (20.5, 54.9) |
| **DEFINE**[22] (RCT with IPD) | Total | 1234 | 394 (31.9) | 38.5 (9.0) | 908 (73.6) | 2.4 (1.2) | 36.9 (17.7, 56.0) |
| | Dimethyl fumarate | 826 | 212 (25.7) | 38.5 (9.0) | 602 (72.9) | 2.4 (1.2) | 36.2 (17.2, 55.1) |
| | Placebo | 408 | 182 (44.6) | 38.5 (9.1) | 306 (75) | 2.5 (1.2) | 38.2 (19.0, 57,5) |
| **Bornstein**[24] (RCT with AD) | Total | 50 | 30 (60.0) | 30.5 (NA) | 29 (58.0) | 3.1 (NA) | 35.6 (19.9, 51.3) |
| | Glatiramer acetate | 25 | 11 (44.0) | 30.0 (NA) | 14 (0.6) | 2.9 (NA) | NA |
| | Placebo | 25 | 19 (76.0) | 31.1 (NA) | 15 (0.6) | 3.2 (NA) | NA |
| **Johnson**[25] (RCT with AD) | Total | 251 | 186 (74.1) | 34.5 (6.4) | 184 (73.3) | 2.6 (1.3) | 30.8 (3.4, 58.1) |
| | Glatiramer acetate | 125 | 89 (71.2) | 34.6 (6.0) | 88 (70.4) | 2.8 (1.2) | NA |
| | Placebo | 126 | 97 (77.0) | 34.3 (6.5) | 96 (76.2) | 2.4 (1.3) | NA |

EDSS: Expanded Disability Status Scale; sd: standard deviation; RCT: Randomized clinical trial; IPD: individual participant data; AD: aggregate data; NA: not available

**Table 2** Estimated parameters from network meta-regression model including the logit-risk as covariate (stage 3).

| Estimated parameters from network meta-regression model | Mean (95% CrI) |
|---|---|
| OR of relapsing for one unit increase in logit-risk ($e^{\gamma_0}$) | 2.72 (2.02, 3.70) |
| OR of relapsing under DF vs placebo ($e^{\delta_{DF}}$) | 0.39 (0.25, 0.59) |
| OR of relapsing under GA vs placebo ($e^{\delta_{GA}}$) | 0.41 (0.22, 0.77) |
| OR of relapsing under N vs placebo ($e^{\delta_N}$) | 0.21 (0.12, 0.34) |
| OR of relapsing under DF vs placebo for one unit increase in logit-risk ($e^{\gamma_{DF}^W}$) | 0.84 (0.44, 1.62) |
| OR of relapsing under GA vs placebo for one unit increase in logit-risk ($e^{\gamma_{GA}^W}$) | 0.64 (0.27, 1.53) |
| OR of relapsing under N vs placebo for one unit increase in logit-risk ($e^{\gamma_N^W}$) | 0.60 (0.30, 1.21) |

DF: Dimethyl fumarate; GA: Glatiramer acetate; N: Natalizumab, CrI.: Credible Interval

Table 3 Predicted % average risk difference and odds ratio of each active treatment versus placebo in the populations in RCTs. Results are shown for all patients, and for two baseline risk groups.

| Treatment effects | Treatment | All patients | Baseline Risk <30% Low-risk patients | Baseline Risk >50% High-risk patients |
|---|---|---|---|---|
| **Risk difference of drug vs placebo (95% CrI)** | Dimethyl fumarate | 38.2 (26.1, 50.4) | 17.6 (10.5, 29.8) | 58.1 (37.9, 73.9) |
| | Glatiramer acetate | 41.1 (25.3, 57.6) | 24.5 (13.8, 42.3) | 56.6 (31.5, 77.0) |
| | Natalizumab | 27.2 (16.1, 40.6) | 15.2 (8.6, 26.2) | 39.0 (20.2, 59.4) |
| **Odds ratio of drug vs placebo (95% CrI)** | Dimethyl fumarate | 0.43 (0.32, 0.57) | 0.50 (0.35, 0.71) | 0.36 (0.25, 0.51) |
| | Glatiramer acetate | 0.53 (0.35, 0.83) | 0.78 (0.49, 1.30) | 0.34 (0.20, 0.61) |
| | Natalizumab | 0.28 (0.20, 0.39) | 0.43 (0.29, 0.62) | 0.17 (0.10, 0.26) |

CrI:Credible Interval;

*Appendix*

When some mean values of the prognostic factors are not reported, we used imputations to allow studies with partial information on covariates to be included in the meta-regression model, as previously described by Hemming et al.[39] This approach models jointly the relationship between the outcome (the average logit of the probability to relapse within the next 2 years - $p_j$) with the covariates and the relations between the covariates, assuming that missing covariates are missing at random. The joint distribution of covariates and outcome can be specified through a series of conditional distributions (assuming continuous covariates):

$$\overline{logit(p_j)}|\overline{PF_{npj}}\dots PF_{1j} \sim N(\mu_{\overline{logit(p_j)}}, \tau_{\overline{logit(p_j)}})$$

$$\overline{PF_{np_J}} | \overline{PF_{(np-1)_J}} \ldots PF_{1_J} \sim N(\mu_{\overline{PF_{np}}}, \tau_{\overline{PF_{np}}})$$

$$\overline{PF_{(np-1)_J}} | \overline{PF_{(np-2)_J}} \ldots \overline{PF_{1_J}} \sim N(\mu_{\overline{PF_{np-1}}}, \tau_{PF_{\overline{np-1}}})$$

$$\vdots$$

$$\overline{PF_{1_J}} \sim N(\mu_{\overline{PF_1}}, \tau_{\overline{PF_1}}),$$

where

$$\mu_{\overline{logit(p_J)}} = \gamma_0 + \gamma_1 \overline{PF_1} + \cdots + \gamma_k \overline{PF_k}$$

$$\mu_{\overline{PF_k}} = \gamma_{k0} + \gamma_{k1} \overline{PF_1} + \cdots + \gamma_{k(k-1)} \overline{PF_{k-1}}$$

$$= \gamma_{k0} + \sum_{i=1}^{k-1} \gamma_{ki} \overline{PF_i}$$

The conditional distribution for the outcome $\overline{logit(p_J)}$ is normal with mean a linear combination of all covariates and a precision parameter $\tau_{\overline{logit(R_J)}}$. Then, each conditional distribution for $\overline{PF_{k_J}}$ is normal with mean a linear combination of the other covariates, $\mu_{\overline{PF_k}}$, and precision parameter $\tau_{PF_k}$. In the case of binary covariates, we used the logit transformation, assuming that it approximately follows a normal distribution.

*Appendix Table 1 Re-calibrated coefficients for the logistic regression prognostic model for the populations in RCTs (stage 2). The estimated coefficients are used to calculate the baseline risk for a new patient in an RCT using Equation 5.*

| Variables | Estimated regression coefficients |
|---|---|
| Intercept | -1.137 |
| Age−37 | -0.025 |
| Disease duration−2.8 | 0.237 |
| EDSS−2.4 | 0.265 |
| Number of gadolinium enhanced lesions (>0 vs 0) | 0.217 |
| Number of previous relapses (1 vs 0) | -0.049 |
| Number of previous relapses (2 or more vs 0) | 0.093 |
| Months since last relapse−2.8 | -0.335 |
| Treatment naïve (Yes vs No) | -0.244 |
| Gender (Female vs Male) | 0.178 |

EDSS: Expanded Disability Status Scale;

Disease duration was transformed to $log$(disease_durration+10)

Months since last realapse was transformed to $log$(months_since_last_relpapse+10)

The coefficients of EDSS, number of gadolinium enhanced lesions and the prior treatment were re-estimated in the "recalibration and selective re-estimation" method.

*Appendix Table 2 Predicted % average risk difference and odds ratio of each active treatment versus placebo in the RCTs population (stage 3), under all three re-calibration methods (stage 2 - Equations 2 to 4). Results are shown for all patients, and for two baseline risk groups.*

| Re-calibration method | Treatment | All patients | Baseline Risk <30% Low-risk patients | Baseline Risk >50% High-risk patients |
|---|---|---|---|---|
| **Re-calibration of intercept (Equation 2)** | Dimethyl fumarate | 31.7 (20.7, 46.3) | 19.3 (10.5, 35.5) | 47.2 (29.0, 66.5) |
| | Glatiramer acetate | 35.6 (21.0, 55.0) | 25.6 (12.6, 48.9) | 47.7 (24.6, 71.7) |
| | Natalizumab | 25.4 (15.2, 40.1) | 18.1 (9.3, 46.7) | 34.4 (18.2, 54.0) |
| **Re-calibration of intercept and overall calibration slope** | Dimethyl fumarate | 31.8 (20.5, 68.1) | 18.7 (10.2, 52.7) | 48.8 (28.3, 91.8) |
| | Glatiramer acetate | 35.4 (20.8, 72.6) | 24.6 (12.2, 37.9) | 48.9 (24.6, 92.9) |
| **(Equation 3)** | Natalizumab | 25.4 (15.1, 58.4) | 17.6 (8.9, 35.6) | 35.5 (17.6, 83.9) |
| **Re-calibration of intercept, overall calibration slope, and re-estimation of some regression coefficients** **(Equation 4)** | Dimethyl fumarate | 38.2 (26.1, 50.4) | 17.6 (10.5, 29.8) | 58.1 (37.9, 73.9) |
| | Glatiramer acetate | 41.1 (25.3, 57.6) | 24.5 (13.8, 42.3) | 56.6 (31.5, 77.0) |
| | Natalizumab | 27.2 (16.1, 40.6) | 15.2 (8.6, 26.2) | 39.0 (20.2, 59.4) |

*Appendix figure 1 Calibration plot of the developed re-calibrated prognostic model with loess smoother, under each re-calibration method (stage 2): A) Re-calibration of intercept methods (via Equation 2), B) Re-calibration of intercept and overall calibration slope method (via Equation 3), C) Re-calibration of intercept, overall calibration slope, and re-estimation of some regression coefficients method (Equation 4). The distribution of the estimated probabilities are shown at the bottom of each graph, by status relapse within two years (i.e. events and non-events). The horizontal axis represents the expected probability of relapsing within 2 years and the vertical axis represents the observed proportion of relapse. The apparent performance measures (c-statistic and c-slope) with their correspondent 95% CI are also shown in the graphs.*

**A) Re-calibration of intercept (Equation 2)**

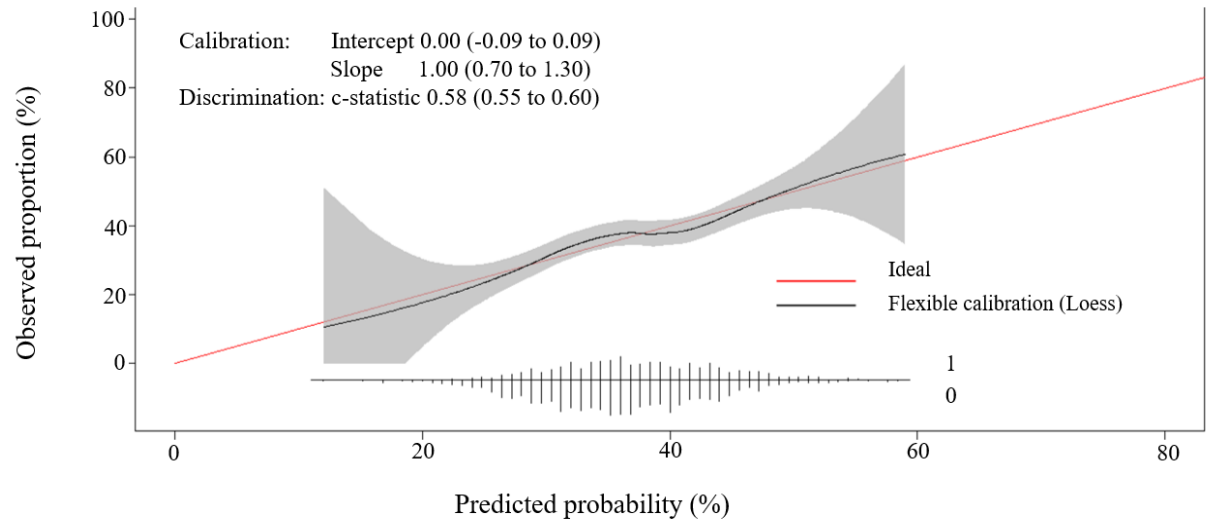

**B) Re-calibration of intercept and overall calibration slope (Equation 3)**

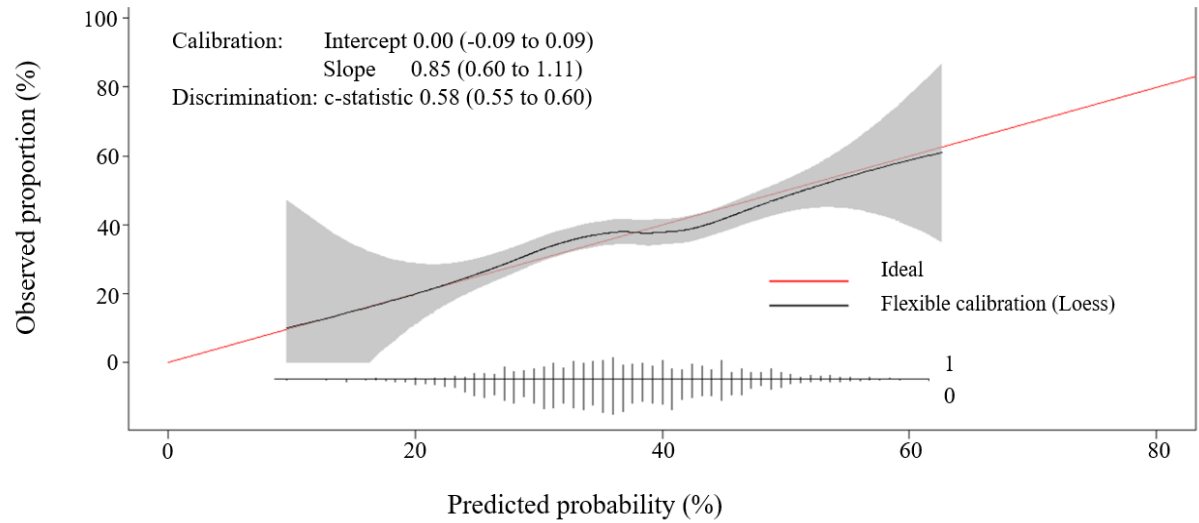

**C) Re-calibration of intercept, overall calibration slope, and re-estimation of some regression coefficients (Equation 4)**

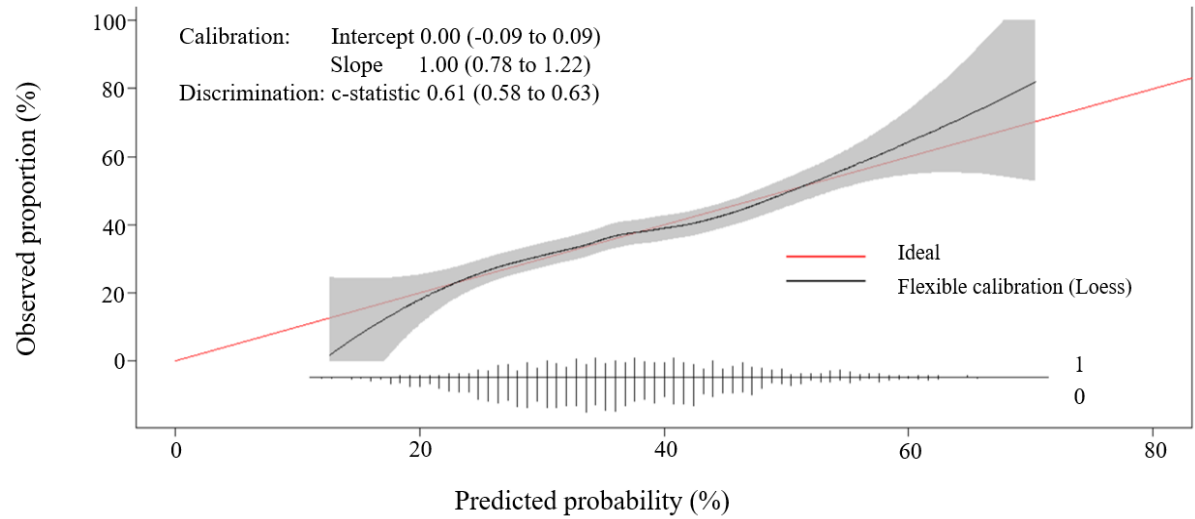

*Appendix figure 2 Predicted probability of relapsing within the next two years as a function of the baseline risk (stage 3) into the randomized clinical trials population, with their 95% CrIs. The x-axis shows the baseline risk of relapsing within the next two years (after re-calibration, stage 2) and the y-axis shows the predicted probability to relapse within the next two years under each one of the available treatments. Between the two red vertical dashed lines are the baseline risk values observed in the three randomized clinical trials with individual participant data.[29,30,31] The distribution of the baseline risk in these three trials is presented at the bottom of the graph.*

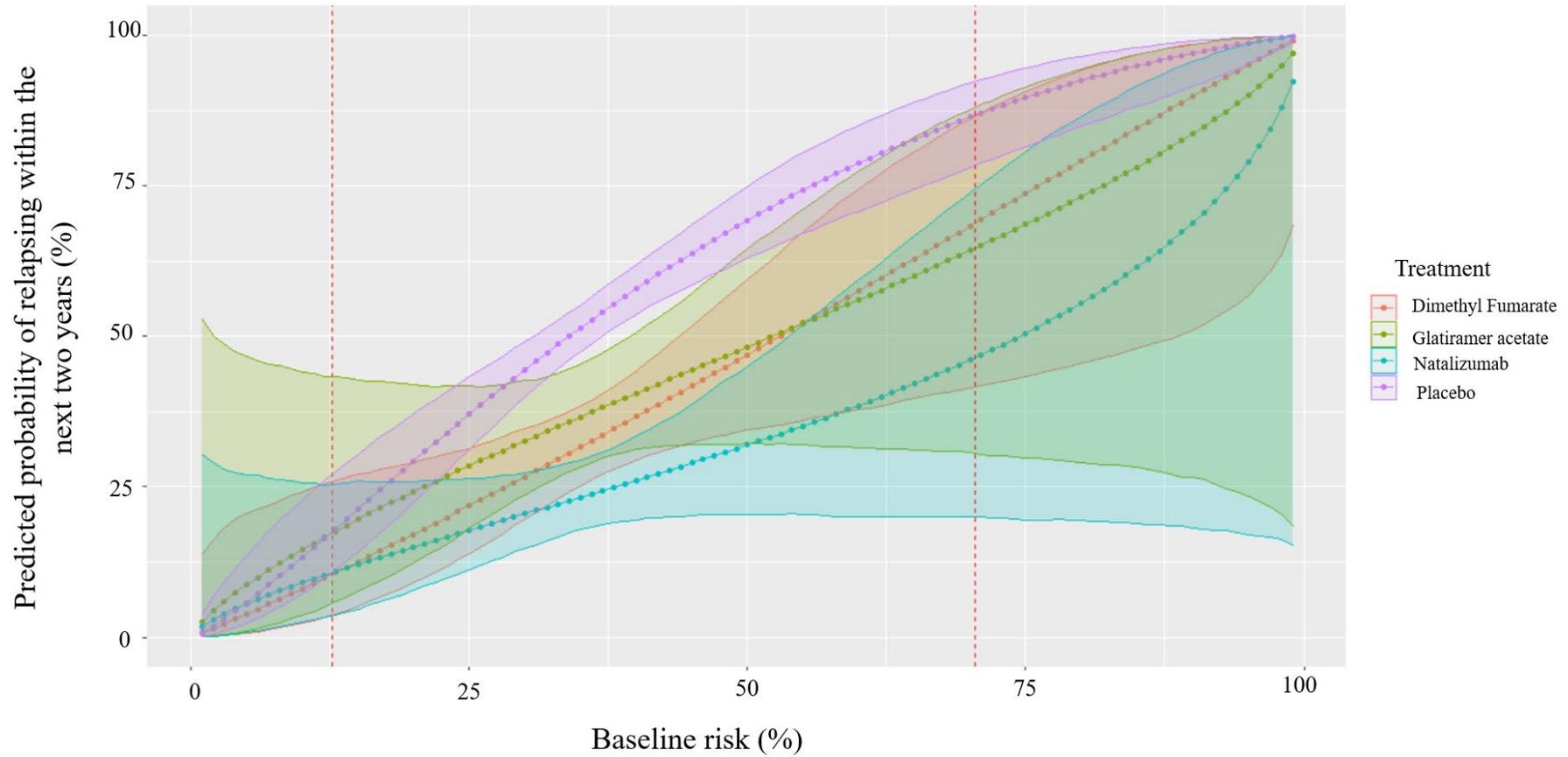

**Appendix figure 4** *Calibration plot of the developed prediction model (stage 3) with loess smoother. The distribution of the estimated probabilities are shown at the bottom of the graph, by status relapse within two years (i.e. events and non-events). The horizontal axis represents the expected probability of relapsing within 2 years and the vertical axis represents the observed proportion of relapse. The apparent performance measures (c-statistic and c-slope) with their correspondent 95% CI are also presented.*

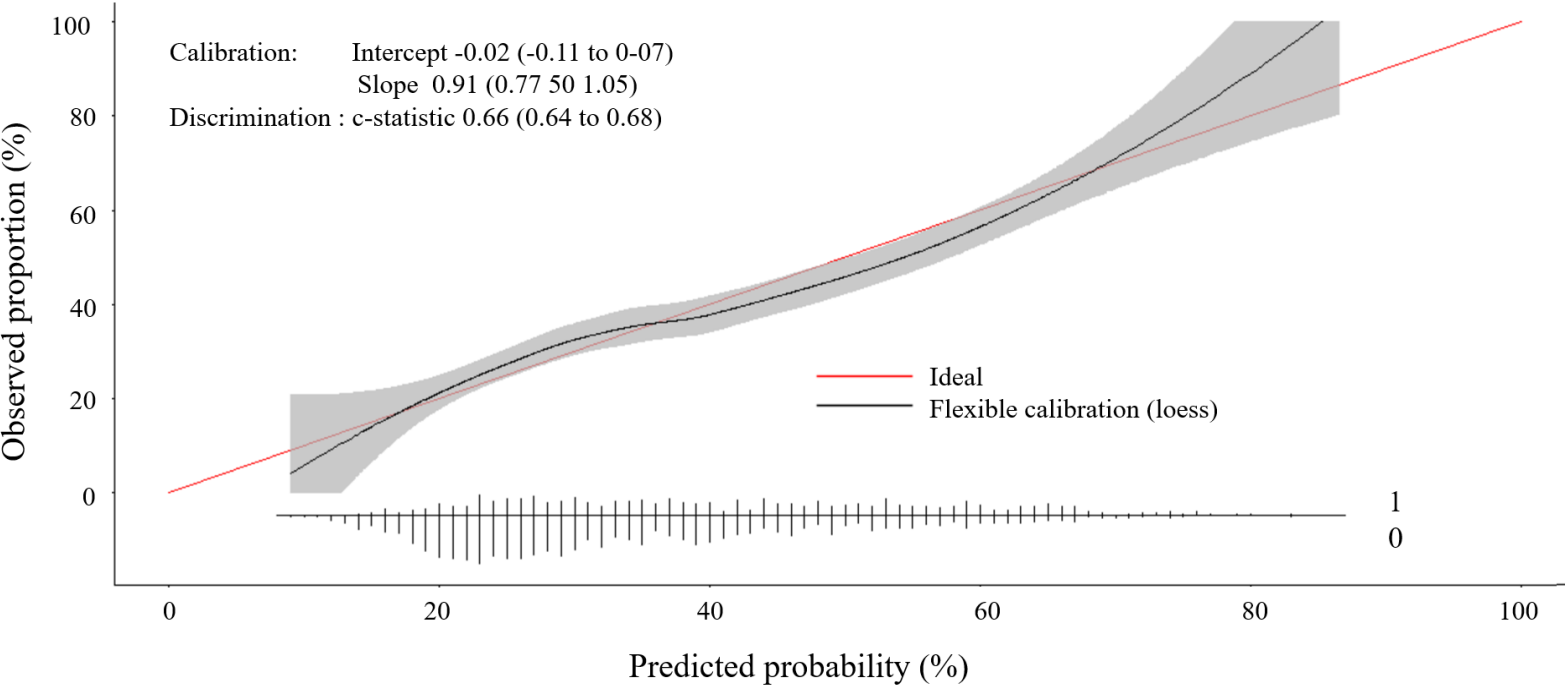

*Appendix figure 4 Predicted % probabilites to relapse within the next two years of each treatment (y-axis) are shown as a function of the baseline risk (x-axis) under all three re-calibration methods (stage 2 - Equations 2 to 4).*

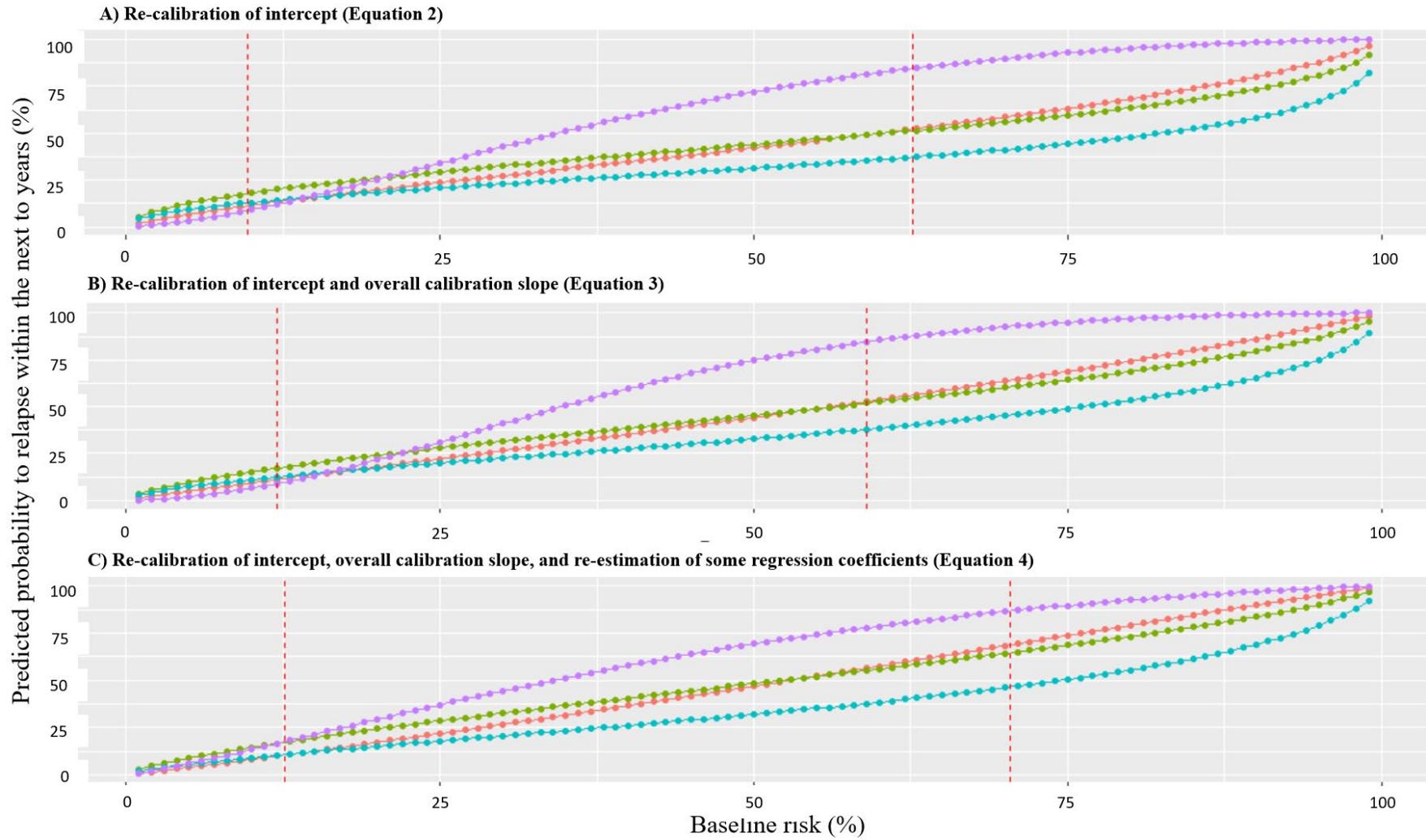

*Appendix figure 5 Predicted probability of relapsing within the next two years as a function of the baseline risk (stage 3) into the Swiss real-world population. The x-axis shows the baseline risk of relapsing within the next two years (after re-calibration, stage 2) and the y-axis shows the predicted probability to relapse within the next two years under each one of the available treatments. Between the two red vertical dashed vertical lines are the baseline risk values observed in Swiss Multiple Sclerosis Cohort (SMSC).[19] The distribution of the baseline risk in this cohort is presented at the bottom of the graph.*

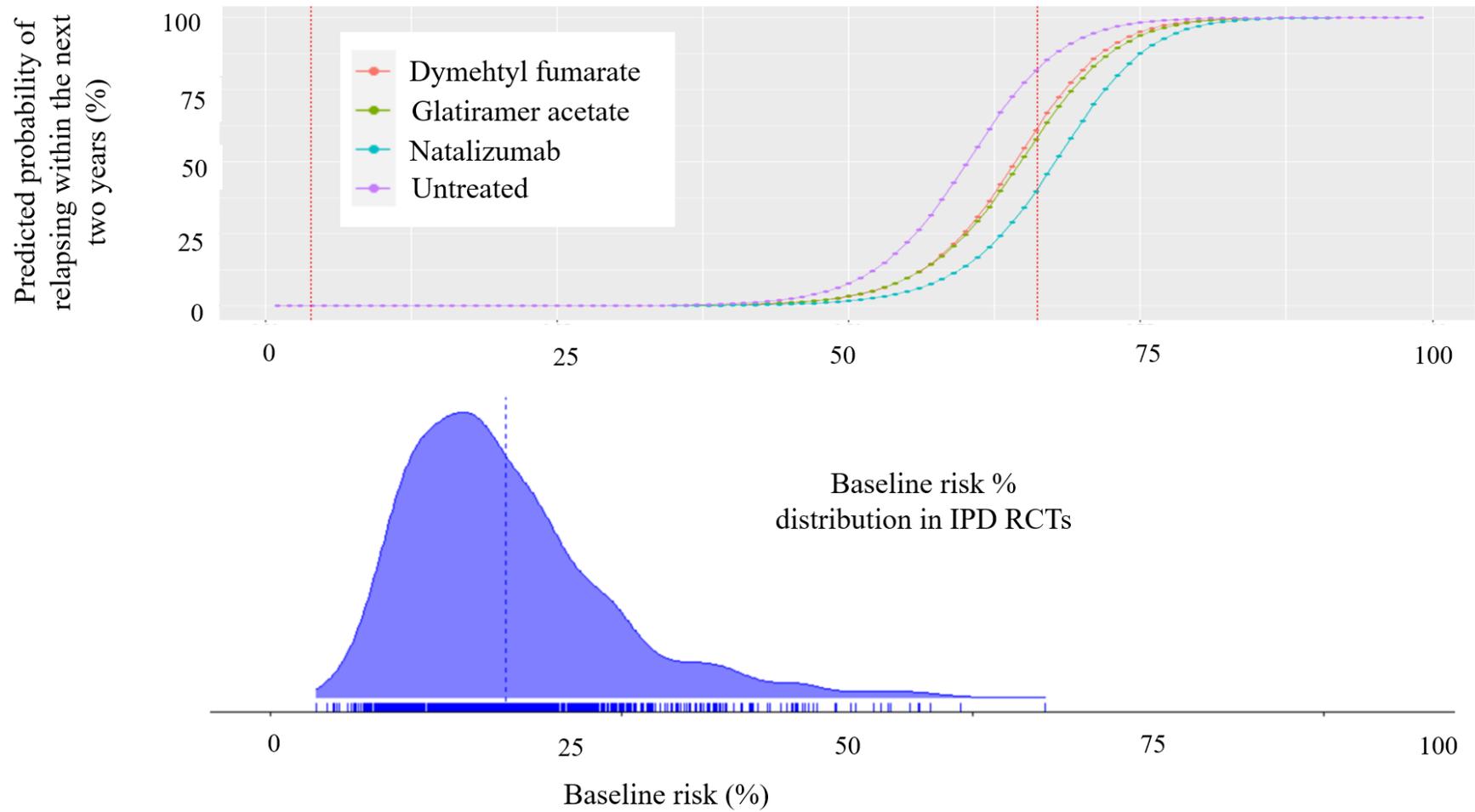